\documentclass[iop]{emulateapj}
\usepackage{amsmath}
\usepackage{color}

\shorttitle{Variability in a Young, L/T Transition Planetary Mass Object}
\shortauthors{Biller et al.}

\begin{document}


\title{Variability in a Young, L/T Transition Planetary Mass Object}


\author{Beth A. Biller\altaffilmark{1,2}, Johanna Vos\altaffilmark{1}, 
Mariangela Bonavita\altaffilmark{1}, 
Esther Buenzli\altaffilmark{2}, Claire Baxter\altaffilmark{1},  
Ian J.M. Crossfield\altaffilmark{3}, 
Katelyn Allers\altaffilmark{4}, 
Michael C. Liu\altaffilmark{5}, 
Micka\"el Bonnefoy\altaffilmark{6}, Niall Deacon\altaffilmark{7}, 
Wolfgang Brandner\altaffilmark{2}, 
Joshua E. Schlieder\altaffilmark{8},  Trent Dupuy\altaffilmark{9}, 
Taisiya Kopytova\altaffilmark{2,10}, Elena Manjavacas\altaffilmark{11},    
France Allard\altaffilmark{12}, 
Derek Homeier\altaffilmark{13},  
Thomas Henning\altaffilmark{2}}

\altaffiltext{1}{Institute for Astronomy, University of Edinburgh, Blackford Hill View, Edinburgh EH9 3HJ, UK}
\altaffiltext{2}{Max-Planck-Institut f\"ur Astronomie, K\"onigstuhl
  17, 69117 Heidelberg, Germany}
\altaffiltext{3}{Lunar \& Planetary Laboratory, University of Arizona,1629 E. University Blvd., Tucson, AZ 95721, USA} 
\altaffiltext{4}{Department of Physics and Astronomy, Bucknell University, Lewisburg, PA 17837, USA}
\altaffiltext{5}{Institute for Astronomy, University of Hawaii, 2680 Woodlawn Drive, Honolulu, HI 96822, USA}
\altaffiltext{6}{University of Grenoble Alpes, IPAG, F-38000 Grenoble, France. CNRS, IPAG, F-38000 Grenoble, France}
\altaffiltext{7}{Centre for Astrophysics Research, University of Hertfordshire, College Lane, Hatfield AL10 9AB, UK}
\altaffiltext{8}{NASA Postdoctoral Program Fellow, NASA Ames Research Center, Moffett Field, CA, USA}
\altaffiltext{9}{Department of Astronomy, University of Texas, 2515 Speedway C1400, Austin, TX 78712}
\altaffiltext{10}{International Max-Planck Research School for Astronomy and Cosmic Physics at the University of Heidelberg, IMPRS-HD,
Germany}
\altaffiltext{11}{Instituto de Astrofísica de Canarias (IAC),  C/ V\'ia L\'actea s/n, E-38205,  La Laguna. Tenerife. Spain}
\altaffiltext{12}{CRAL-ENS, 46, All\'ee d'Italie, 69364 Lyon Cedex 07 France}
\altaffiltext{13}{Zentrum f{\"u}r Astronomie der Universit{\"a}t
  Heidelberg, Landessternwarte, K\"onigstuhl 12, 69117 Heidelberg, Germany}







\begin{abstract}
  As part of our ongoing NTT SoFI survey for variability in young
  free-floating planets and low mass brown dwarfs, we detect
  significant variability in the young, free-floating planetary mass
  object PSO J318.5-22, likely due to rotational modulation of
  inhomogeneous cloud cover. A member of the 23$\pm$3 Myr $\beta$ 
Pic moving group,  PSO J318.5-22 has 
T$_\mathrm{eff}$ = 1160$^{+30}_{-40}$ K and a mass estimate of
8.3$\pm$0.5 M$_{Jup}$ for 
a 23$\pm$3 Myr age.  PSO J318.5-22 is intermediate
in mass between 51 Eri b and $\beta$ Pic b, the two known exoplanet companions in the $\beta$ Pic 
moving group.  With variability amplitudes from
  7-10$\%$ in J$_{S}$ at two separate epochs over 3-5 hour
  observations, we constrain the rotational period of this object 
to $>$5 hours.  In K$_{S}$, we
  marginally detect a variability trend of up to 3$\%$ over a 3 hour
  observation.  
This is the first detection of weather on an extrasolar planetary
  mass object.  Among L dwarfs surveyed at high-photometric precision
  ($<$3$\%$) this is the highest
  amplitude variability detection.  Given the low surface gravity of this
  object, the high amplitude preliminarily suggests that such objects may 
be more variable than their high mass counterparts, although 
observations of a larger sample is necessary to confirm this.
Measuring similar variability for directly imaged planetary companions 
is possible with instruments such as SPHERE and GPI and will provide 
important constraints on formation.  Measuring variability at 
multiple wavelengths can help constrain cloud structure.
\end{abstract}
\footnote{Based on observations made with ESO Telescopes at the La
  Silla Paranal Observatory under programme ID 095.C-0590}





\section{Introduction}

Of the current ensemble of $\sim$30 free-floating young
planetary mass objects \citep[]{Gag14, Gag15}, PSO J318.5-22
\citep[]{Liu13} is the closest analogue in properties to imaged
exoplanet companions.  \citet[]{Gag14} and
\citet[]{Liu13} identify it as a $\beta$ Pic moving group member 
\citep[23$\pm$3~Myr,][]{Mam14}
and it possesses 
colors and magnitudes similar to the HR 8799
planets \citep[][]{Mar08,Mar10} and 2M1207-39b \citep[][]{Cha05}. 
PSO J318.5-22 has 
T$_\mathrm{eff}$ = 1160$^{+30}_{-40}$ K and a published mass estimate of
6.5$^{+1.3}_{-1.0}$ M$_{Jup}$ for an age of 12 Myr \citep[]{Liu13}, 
rising to 8.3$\pm$0.5 M$_{Jup}$ for the updated age of 23$\pm$3 Myr 
(Allers et al. submitted).   PSO J318.5-22 is intermediate
in mass and luminosity between 51 Eri b \citep[$\sim$2~M$_{Jup}$,][]{Mac15} and $\beta$ Pic b 
\citep[$\sim$11-12~M$_{Jup}$,][]{Lag10, Bon14}, the two known exoplanet companions in the $\beta$ Pic 
moving group.
Because PSO J318.5-22 is free-floating, it enables high
precision characterization not currently possible for exoplanet companions to bright
stars.  In particular, we report here the first detection of 
photometric variability in a young, L/T transition planetary mass object.

Variability is common for cool brown dwarfs
but until now has not been probed for lower-mass planetary objects
with similar effective temperatures. 
Recent large-scale surveys of brown dwarf variability
with Spitzer have revealed mid-IR variability of up to a few percent
in $>$50\% of L and T type brown dwarfs \citep[]{Met15}.
\citet[]{Bue14} find that $\sim$30\% of the L5-T6 objects surveyed in
their HST SNAP survey show variability trends and large ground-based
surveys also find widespread variability \citep[]{Rad14a, Wil14, Rad14b}.
While variability amplitude may be increased across the L/T transition
\citep[]{Rad14a}, variability is now robustly observed
across a wide range of L and T spectral types.  We therefore
expect variability in young extrasolar planets, which share
similar T$_\mathrm{eff}$ and spectral types but lower surface gravity.
In fact, 
\citet[]{Met15} tentatively find a correlation between low surface gravity and 
high-amplitude variability in their L dwarf sample.

Observed field brown dwarf variability is likely produced by rotational
modulation of inhomogenous cloud cover over the 3-12 hour rotational periods
of these objects \citep[]{Zap06}.
\citet[]{Apa13} and \citet[]{Bue15} find that their
variability amplitude as a function of wavelength
are best fit by a combination of thin and thick cloud
layers.  We expect a similar mechanism to drive variability in
planetary mass objects with similar T$_\mathrm{eff}$, albeit with
potentially longer periods, as these objects will not yet have 
spun up with age.  Only a handful of
directly imaged exoplanet companions are amenable to variability searches
using high-contrast imagers such as SPHERE at the VLT \citep[]{Beu08} and GPI at
Gemini \citep[]{Mac14}; to search for variability in a larger sample of planetary mass
objects and young, very low mass brown  dwarfs, 
we have been conducting the first survey for free-floating
planet variability using NTT SoFI \citep[]{Moo98}.  We have observed 22 objects to
date, of which 7 have mass estimates $<$13 M$_{Jup}$ and all have mass
estimates $<$25 M$_{Jup}$.  PSO J318.5-22 is the first variability
detection from this survey.  

\section{Observations and Data Reduction}

\begin{deluxetable}{lccccc}
\tablecolumns{10}
\tablewidth{3in}
\tabletypesize{\tiny}
\tablecaption{SOFI observations of PSO J318.5-22\label{tab:obs}}
\tablehead{
\colhead{Date} & \colhead{Filter}  & \colhead{DIT} &
\colhead{NDIT} & \colhead{Exp. Time} & \colhead{On-Sky Time}} 
\startdata
2014 Oct 9 & J$_{S}$ & 10 s & 6 & 3.80 hours & 5.15 hours \\
2014 Nov 9 & J$_{S}$ & 15 s & 6 & 2.40 hours & 2.83 hours \\
2014 Nov 10 & K$_{S}$ & 20 s & 6 & 2.80 hours & 3.16 hours \\
\enddata 
\end{deluxetable}

We obtained 3 datasets for PSO J318.5-22 with NTT SoFI 
(0.288$\arcsec$/ pixel, 4.92'$\times$4.92' field of view) 
in October and November 2014.
Observations are presented in Table~\ref{tab:obs}.
We attempted to cover as much of the unknown rotation period
as possible, however, scheduling constraints and weather conditions
limited our observations to 2-5 hours on sky.  In search mode, we 
observed in  J$_{S}$, however we did obtain a K$_{S}$ followup
lightcurve for PSO J318.5-22.  
We nodded the target between two positions on the chip, ensuring that, at each jump from
position to position, the object is accurately placed on the same
original pixel.  This allowed for sky-subtraction, while 
preserving photometric stability.  We followed an ABBA
nodding pattern, taking three exposures at each nod position.

Data were corrected for crosstalk artifacts between quadrants, flat-fielded using special 
dome flats which correct for the ``shade'' (illumination dependent 
bias) found in SoFI images, and illumination corrected using
observations of a standard star.
Sky frames for each nod position were created by median 
combining normalized frames from the other nod positions 
closest in time. These were then re-scaled to and subtracted
from the science frame.  Aperture photometry 
for all sources on the frame were acquired using
the IDL task aper.pro with aperture radii of 4, 4.5, 5, 5.5, 6, and 6.5 pixels and
background subtraction annuli from 21-31 pixels.  


\section{Light Curves}

\begin{figure}
\includegraphics[width=3.5in]{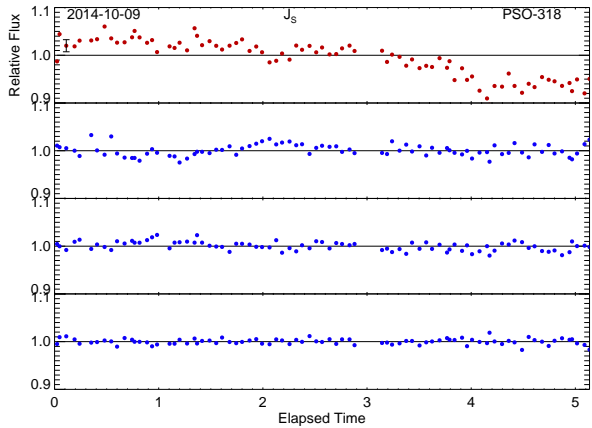}
\caption{ Final binned J$_S$ lightcurve and comparison
detrended reference stars from 9 October 2014.
Typical error bars are shown 
on the first photometric point. 
The variability amplitude at this epoch is $>$10$\%$ with a period 
of $>$5 hours.  \label{fig:lightcurves1}
}
\end{figure}

\begin{figure}
\includegraphics[width=3.5in]{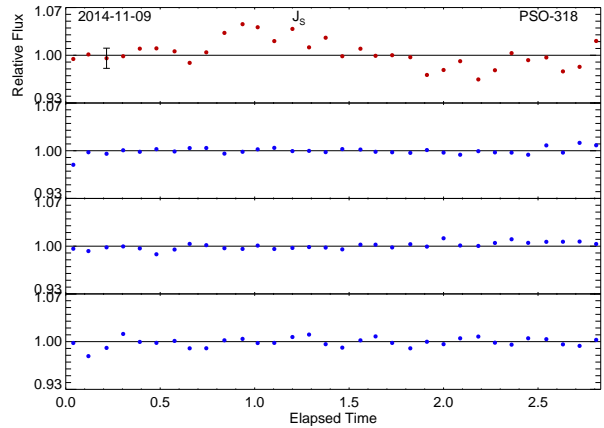} 
\includegraphics[width=3.5in]{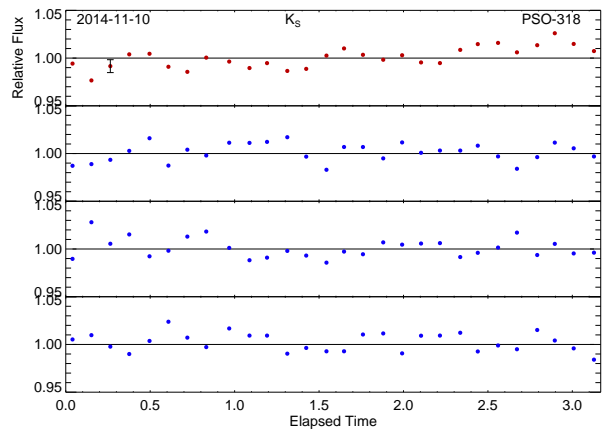} 
\caption{
Top: Final binned J$_S$ lightcurve from 9 November 2014.
Bottom: Final binned K$_{S}$ lightcurve from 11 November 2014.
Lightcurves are presented similarly as in Fig.~\ref{fig:lightcurves1}.
The
J$_{S}$ variability amplitude at this epoch is $>$7$\%$ with a period 
of $\geq$3 hours.  We marginally detect K$_{S}$ variability,
with amplitude up to 3$\%$ over our 3 hour observation.
\label{fig:lightcurves2}}
\end{figure}

We present the final binned  J$_{S}$ lightcurve from October 2014 
(with detrended reference stars for comparison) in 
Fig.~\ref{fig:lightcurves1} and the final binned J$_{S}$ and K$_{S}$
lightcurves from November
2014 in Fig.~\ref{fig:lightcurves2}.  Raw light curves
obtained from aperture photometry display fluctuations in brightness
due to changing atmospheric transparency, airmass, and residual
instrumental effects.  
These changes 
can be removed via
division of a calibration curve calculated from carefully
chosen, well-behaved reference stars \citep[]{Rad14a}.
To detrend our lightcurves, first we discarded
potential reference stars with peak flux values below 10 or greater
than 10000 ADU (where array non-linearity is limited to $<$1.5$\%$). Different nods
were normalized via division by their median flux before being
combined to give a relative flux light curve. For each star a
calibration curve was created by median combining all other reference
stars (excluding that of the target and star in question). The
standard deviation and linear slope for each lightcurve was calculated
and stars with a standard deviation or slope $\sim$1.5-3 times greater
than that of the target were discarded. This process was iterated 
until a set of well-behaved reference stars was
chosen. Final detrended light curves were obtained by dividing the raw
curve for each star by its calibration curve.  The best lightcurves
shown here are with the aperture that minimizes the standard deviation
after removing a smooth polynomial \citep[as~done~in~][]{Bil13} --
for all epochs, the 4 pixel aperture (similar to the PSF FWHM) yielded
the best result.
Final lightcurves 
are shown binned by a factor of three -- combining all three exposures
taken in each ABBA nod position.  Error bars were calculated in 
a similar manner as in \citet[]{Bil13} -- a low-order polynomial was fit to 
the final lightcurve and then subtracted to remove any astrophysical 
variability and the standard deviation of the subtracted lightcurve was
adopted as the typical error on a given photometric point 
(shown in each lightcurve as the error bar given on the first
photometric point).  As a check, we also
measured photometry and light curves using both the publically
available aperture photometry pipeline from \citet[]{Rad14b} as well as
the psf-fitting pipeline described in \citet[]{Bil13}.  Results
from all three pipelines were consistent.

We found the highest amplitude of variability in our J$_{S}$ 
lightcurve from 9 October 2014 -- over the
five hours observed, PSO J318.5-22 varies by 10$\pm$1.3$\%$.
The observed variability does not correlate with airmass 
changes -- the target was overhead for the majority of this observation,
with airmass between 1 and 1.2 for the first
3 hours, increasing to $\sim$2 by the end of the observation. 
The flattening of the lightcurve from 4-5 hours elapsed time in our
lightcurve may be indicative of a minimum in the lightcurve.  
However, as no clear repetition of maxima or minima have been covered,
the strongest constraints we can place on the rotational period and variability
amplitude for PSO J318.5-22 in this epoch is that the period
must be $>$5 hours and the amplitude must be $\geq$10$\%$.
If the variation is sinusoidal, these observations point to an 
even longer period of $>$7-8 hours.

On 9 November 2014, we recovered J$_{S}$ variability with a 
somewhat smaller amplitude of 7$\pm$1$\%$ over our three hour long
observation.  A maximum is seen 1 hour into the observation
and a potential minimum is seen at 2 hours into the observation.
The observed variability is not correlated with airmass changes 
during the observation -- the observation started at
airmass = 1.1, increasing steadily to airmass =2.0 at the end of the
observation.
If the variability is roughly sinusoidal and single peaked, 
this observation would suggest a period$\sim$3 hours; however, 
we cannot constrain the period beyond requiring it to be  
$>$3 hours, as we have not covered 
multiple extrema and as the light curve could potentially be double-peaked
\citep[]{Rad12}.  The lightcurve evolved considerably
between the October and November 2014 epochs -- a phenomena also found  
in other older variable brown dwarfs \citep[]{Rad12, Rad14a, Art09,
  Met15,Gil13}.

On 11 November 2014, we obtained a K$_{S}$ lightcurve for PSO J318.5-22.
Given its extremely red colors, PSO J318.5-22 is brighter
in K$_{S}$ than J$_{S}$ and is one of the brightest objects in the
SoFI field.  Thus, we attain higher photometric
precision in our K$_{S}$ (0.7$\%$) lightcurve compared to J$_{S}$ (1 -
1.3$\%$).  Fitting slopes to the target and 3 
similarly-bright reference stars, the target increases in flux by 0.9$\%$ per hour 
while the reference stars have slopes of 0.1-0.6$\%$ / hour
(consistent with a flat line within our photometric precision).
Thus, we tentatively find a marginal variability trend of up to 3$\%$ over
our 3 hour observation, requiring reobservation to be confirmed.
Additionally, in this case 
the tentative variability is not completely uncorrelated with airmass
changes -- during this observation, airmass increased steadily from 1.1 to 
2.2.  

\section{Discussion}

This is the first detection of variability in such a cool, low-surface
gravity object.  While variability has been 
detected previously for very young ($<$1-2 Myr) planetary mass
objects in star-forming regions such as Orion \citep[cf.~][]{Joe13}, 
such variability is driven by a different mechanism than 
expected for PSO J318.5-22.  These previous detections have been for 
M spectral type objects with much higher T$_\mathrm{eff}$ than PSO J318.5-22. 
At these temperatures, variability is driven by starspots induced by 
the magnetic fields of these objects or ongoing accretion.  
PSO J318.5-22 is too cool to have starspots and 
likely too old for ongoing accretion.  From its red colors, PSO
J318.5-22 must be entirely cloudy \citep[]{Liu13}.  Thus the likely mechanism producing 
the observed variability is inhomogeneous cloud cover, as has
been found previously to drive variability in higher mass brown dwarfs
with similar T$_\mathrm{eff}$ \citep[]{Art09, Rad12, Bue14, Rad14a,
  Rad14b, Wil14, Apa13, Bue15}.  Notably, among 
L dwarfs surveyed at high-photometric precision  ($<$3$\%$),
PSO J318.5-22's J band
variability amplitude is the highest measured for an L dwarf to date
(cf. Yang et al. 2015 and Buenzli et al. submitted) -- reinforcing
the suggestion by \citet[]{Met15} 
 that variability amplitudes might be typically larger for lower
 gravity objects. 

To model cloud-driven as well as hot-spot variability, 
we follow the approach of \citet[]{Art09} and \citet[]{Rad12}, 
combining multiple 1-d models to represent different regions of cloud cover.
We consider the observed atmosphere of our object to be composed of
flux from two distinct cloud regions (varying in temperature and/or in cloud 
prescription) with fluxes of $F_{1}$ and $F_{2}$ respectively and with 
a minimum filling fraction for the $F_{2}$ region of $a$.
The peak-to-trough amplitude of variability ($\Delta F$ / $F$, i.e. 
the change of flux divided by the mid-brightness flux) observed in a given 
bandpass due a change of filling fraction over the course of the
observation is given by Equation 2 from Radigan et al. 2012, 
where $\Delta$a is the change in filling factor over the observation, 
$\Delta$F = $F_{2}$ - $F_{1}$, and $\alpha$ = $a$ + 0.5$\Delta a$, 
the filling fraction of the $F_{2}$ regions at mid-brightness:

\begin{equation}
\begin{array}{l}
A = \frac{(1-a-\Delta a) F_{1} + (a + \Delta a) F_{2} - (1 - a) F_{1}
  - a F_{2}}{0.5[(1-a-\Delta a) F_{1} + (a + \Delta a) F_{2} + (1 - a) F_{1}
  + a F_{2}]} \\
 =\frac{\Delta a }{\alpha + F_{1} / \Delta F} \\
\end{array}
\label{eq:ff}
\end{equation}

We calculated the synthetic photon fluxes $F_{1}$ and $F_{2}$ using
the cloudy exoplanet models of \citet[]{Mad11} and the filter
transmissions provided for the SoFI $J_{S}$ and $K_{S}$ filters.  
While a diversity of brown dwarf / exoplanet cloud models are 
available \citep[e.g.~][]{Sau08, All01, All12}, the \citet[]{Mad11}
models are particularly tuned to fit the
cloudy atmospheres and extremely 
red colors of young low-surface gravity objects such as the HR 8799
exoplanets \citep[]{Mar08, Mar10}.  As PSO J318.5-22 is a
free-floating analogue of these exoplanets, the \citet[]{Mad11}
models are the optimal choice for this analysis. 
Because PSO J318.5-22's extraordinarily red colors preclude clear
patches in its atmosphere \citep[]{Liu13}, we consider
only combinations of cloudy models.  
The \citet[]{Mad11} models model the cloud distribution
according to a shape function, $f(P)$:

\begin{equation}
f(P) = 
\begin{cases}
(P/P_{\rm u})^{s_{\rm u}} &  P \le P_{\rm u} \\
f_{\rm    cloud} &  P_{\rm u} \le P \le P_{\rm d} \\ 
\left(P/P_{\rm d}\right)^{-s_{\rm d}}&   P \ge P_{\rm d} \, , \\
\end{cases}
\label{eq:cloud}
\end{equation} 

where P$_{u}$ and P$_{d}$ are the pressures at the upper and lower pressure
cutoffs of the cloud and P$_{u}$ $<$ P$_{d}$.  The indices
s$_{u}$ and s$_{d}$ control how rapidly the clouds dissipate at their
upper and lower boundaries. We consider combinations of 3
cloud models from \citet[]{Mad11}, with 60 $\mu$m 
grain sizes and solar metallicity:  
\begin{equation}
\begin{array}{l}

\mbox{Model E: } s_{\rm u}=6, s_{\rm d}=10, f_{\rm cloud}=1 \\

\mbox{Model A: } s_{\rm u}=0, s_{\rm d}=10, f_{\rm cloud}=1 \\

\mbox{Model AE: } s_{\rm u}=1, s_{\rm d}=10, f_{\rm cloud}=1 \\
\end{array}
\end{equation}

where model E cuts off rapidly at altitude, model A provides the
thickest clouds, extending all the way to the top of the atmosphere,
and model AE provides an intermediate case.  

\begin{figure}
\begin{tabular}{c}
\includegraphics[width=3.5in]{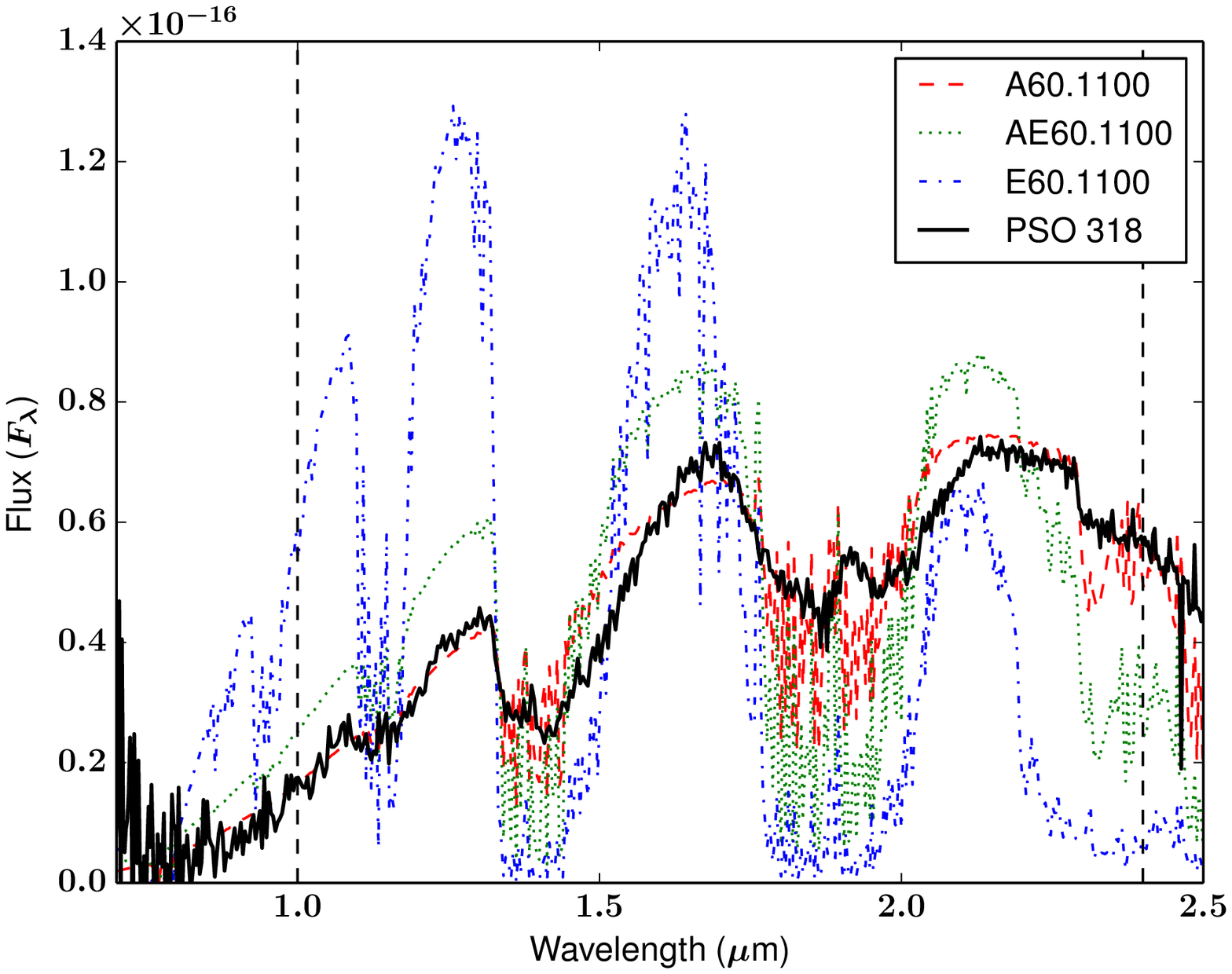} \\
\includegraphics[width=3.5in]{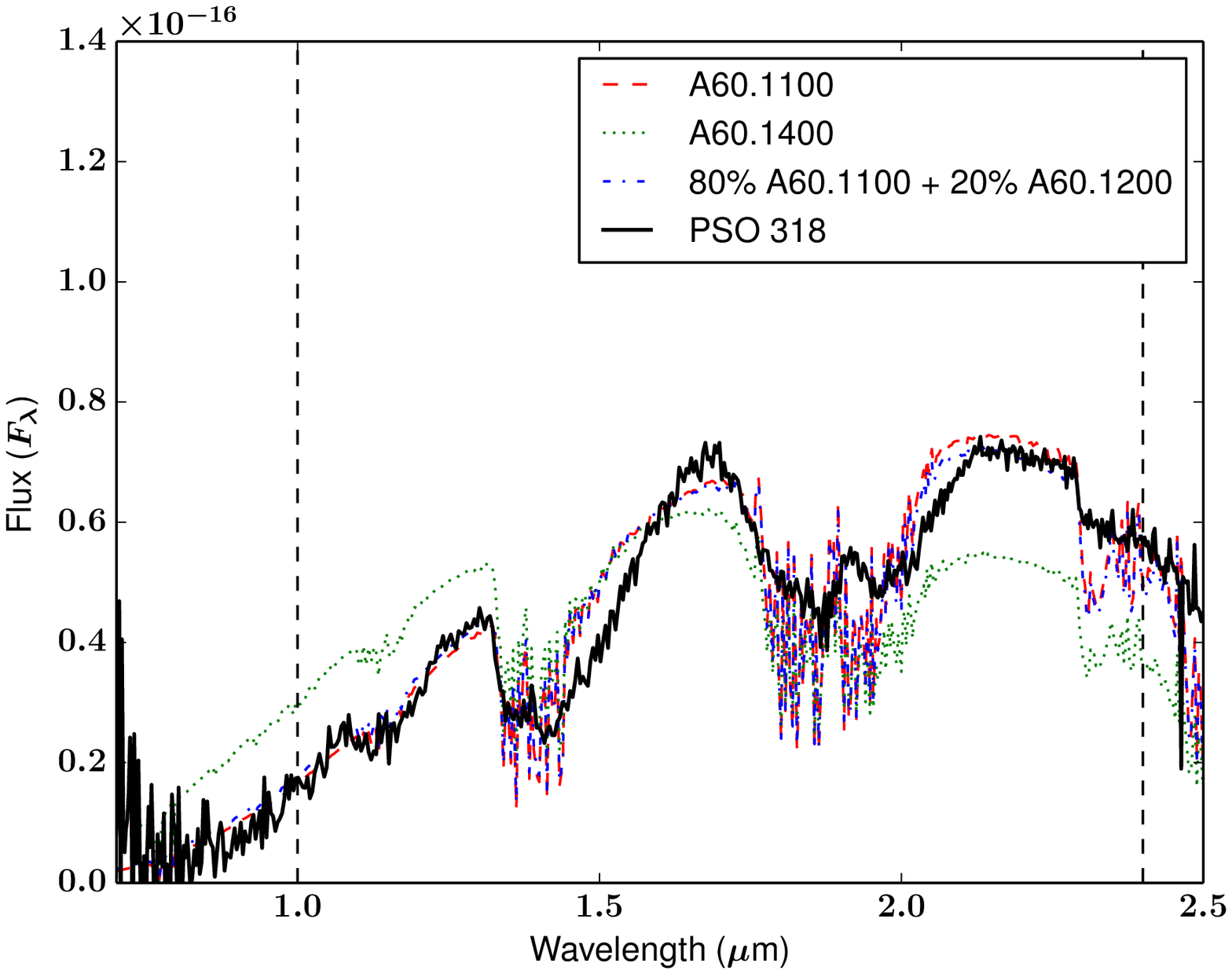}\\
\end{tabular}
\caption{Top: Best fit single model spectra for 
model A, AE, and E clouds from \citet[]{Mad11} 
overlaid on the spectrum of PSO J318.5-22 presented in \citet[]{Liu13}.  
The overall best fit model has
T$_{\mathrm{eff}}$=1100 K, log(g)=4, solar metallicity, and model A (thick) clouds.  
AE and E cloud models fail to reproduce the observed spectrum.
Bottomt: T$_{\mathrm{eff}}$=1400 K, log(g)=4, solar metallicity,
and model A (thick) cloud spectrum as well as the
best fit multi-component spectrum, consisting of 80$\%$ T$_{\mathrm{eff}}$=1100 K + 20$\%$ 
 T$_{\mathrm{eff}}$=1200 K model A clouds overplotted on the
 \citet[]{Liu13} spectrum.  Hotter models do not fit the observed
 features of the \citet[]{Liu13} spectrum; the combined 1100 K
 + 1200 K model spectrum reproduces the observed spectrum marginally better
 than the best fit single component model.
\label{fig:spectrum}}
\end{figure}

Fitting single component models to the spectrum presented in \citet[]{Liu13}, 
we find that the best single component fit is for A prescription clouds
with  T$_\mathrm{eff}$=1100 K (see Fig.~\ref{fig:spectrum}).   This agrees well
with the derived T$_\mathrm{eff}$=1100$^{+30}_{-40}$ K from \citet[]{Liu13}.
We thus adopt T$_\mathrm{eff}$=1100 K as the temperature of the dominant 
cloud component, with a second cloud component at T$_{2}$.  Explicitly
fitting multi-cloud component models, we find that a combination of
80$\%$ model A clouds with T$_\mathrm{eff}$=1100 K and 20$\%$ 
model A clouds with T$_\mathrm{eff}$=1200 K 
marginally fit the spectrum better than a single component fit.  Multi-component
fits using multiple cloud prescriptions do not fit the spectrum well
-- model A clouds (or similar) are likely
the dominant cloud component in this atmosphere.
We did not attempt further analysis of the spectrum in terms of variable
cloud components, as the spectrum was observed at a different epoch 
than the variability monitoring.  

\begin{figure}
\begin{tabular}{cc}
\includegraphics[width=3in]{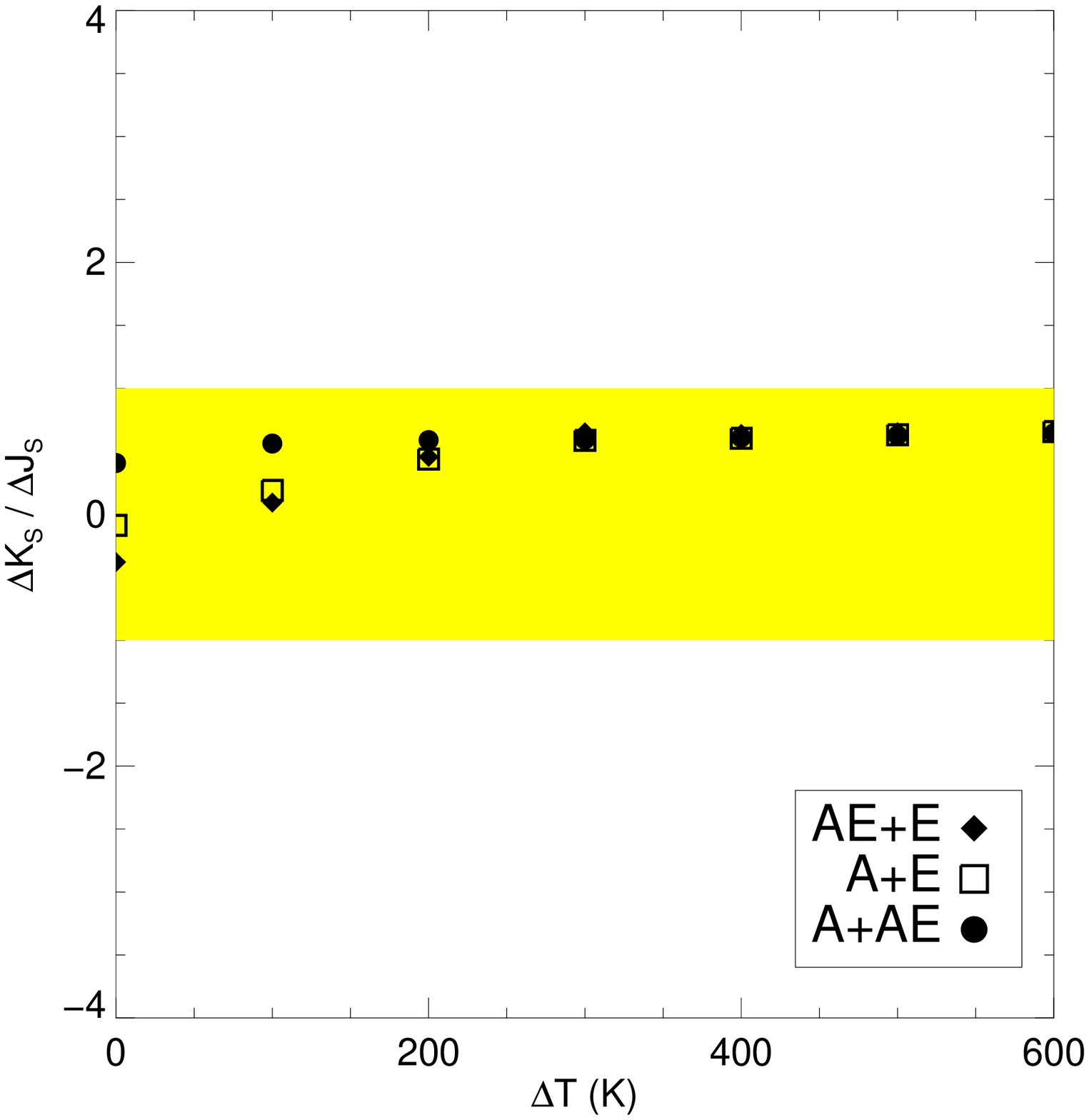} &
\includegraphics[width=3in]{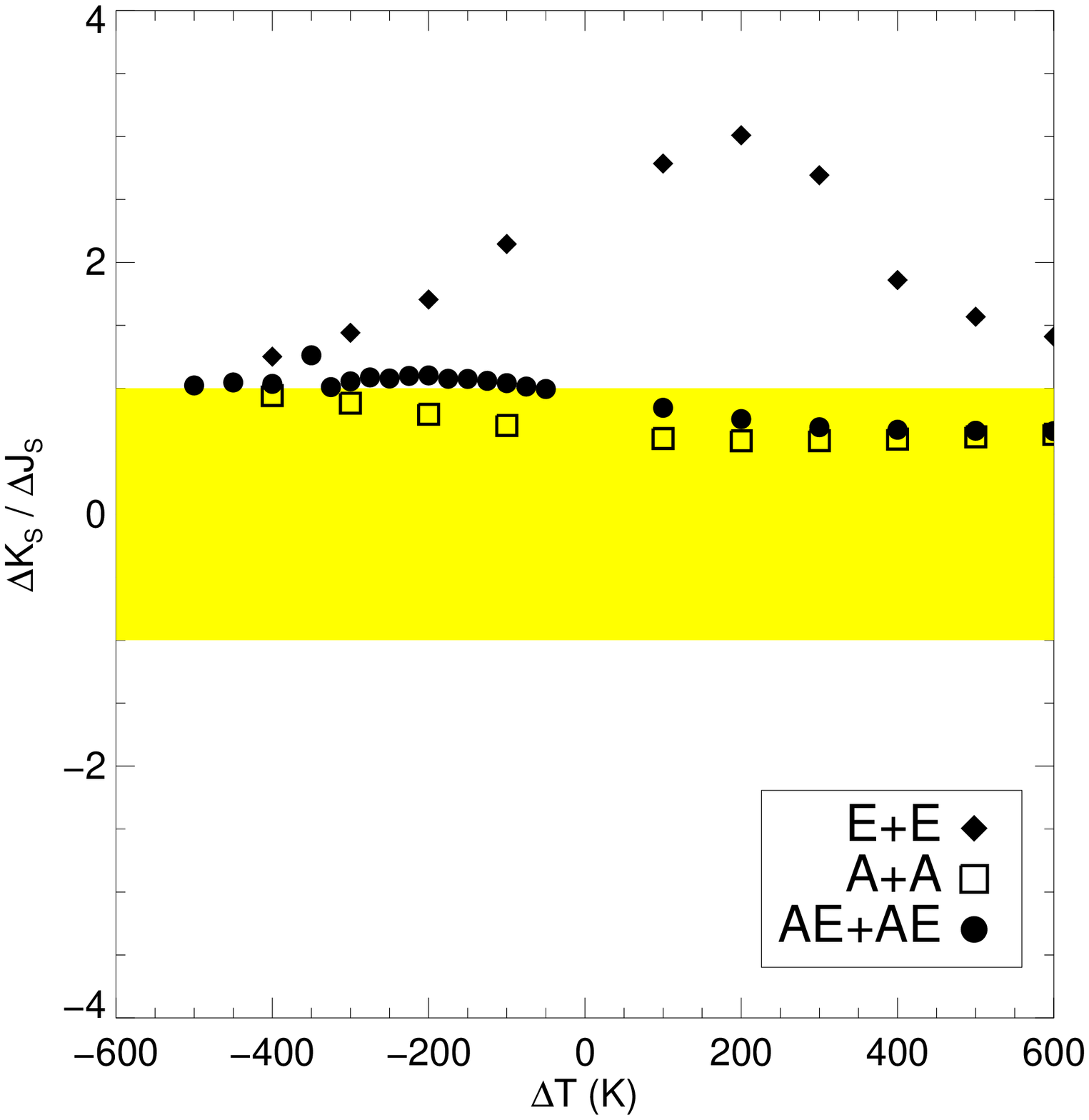} \\
\includegraphics[width=3in]{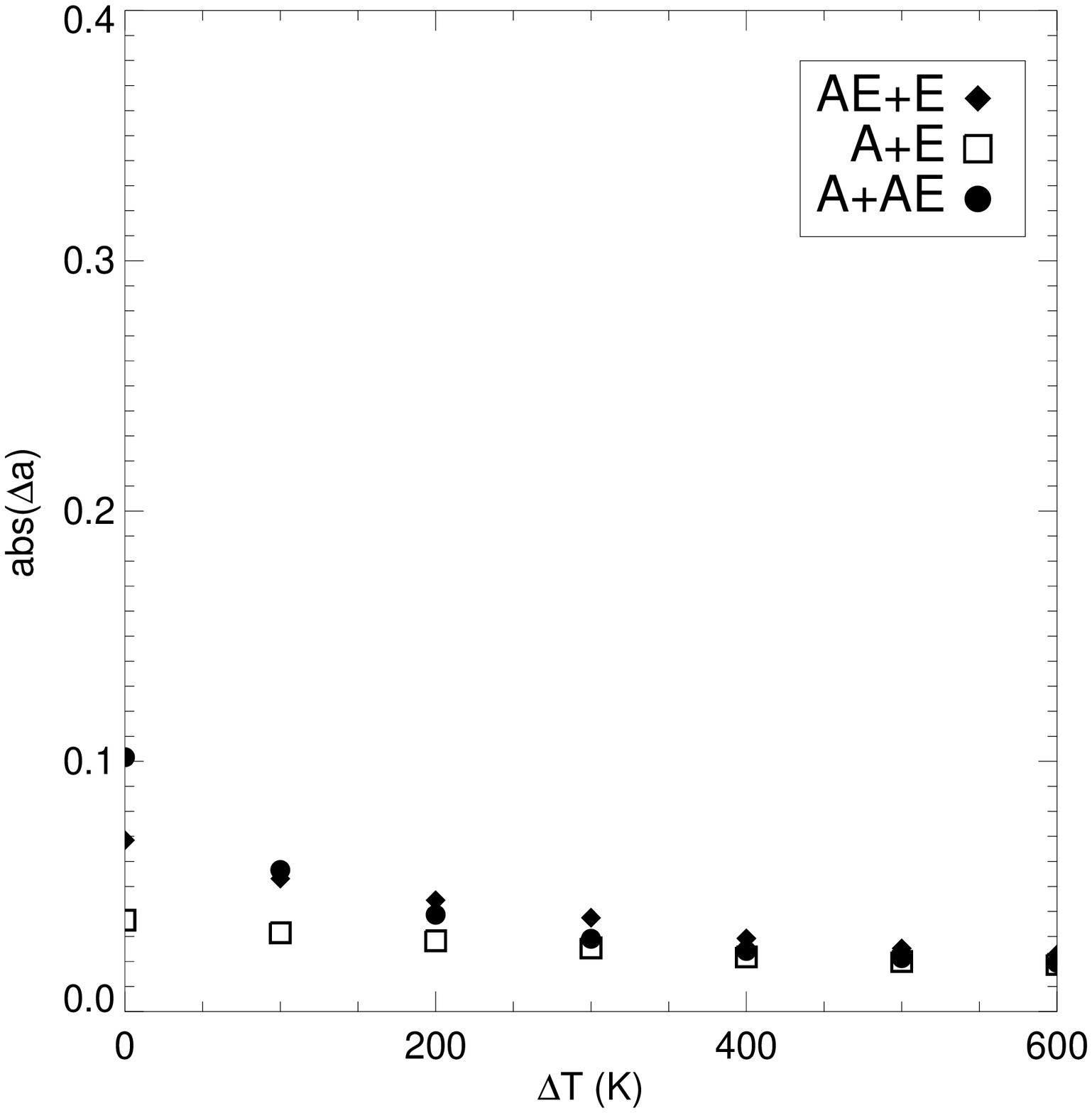} & 
\includegraphics[width=3in]{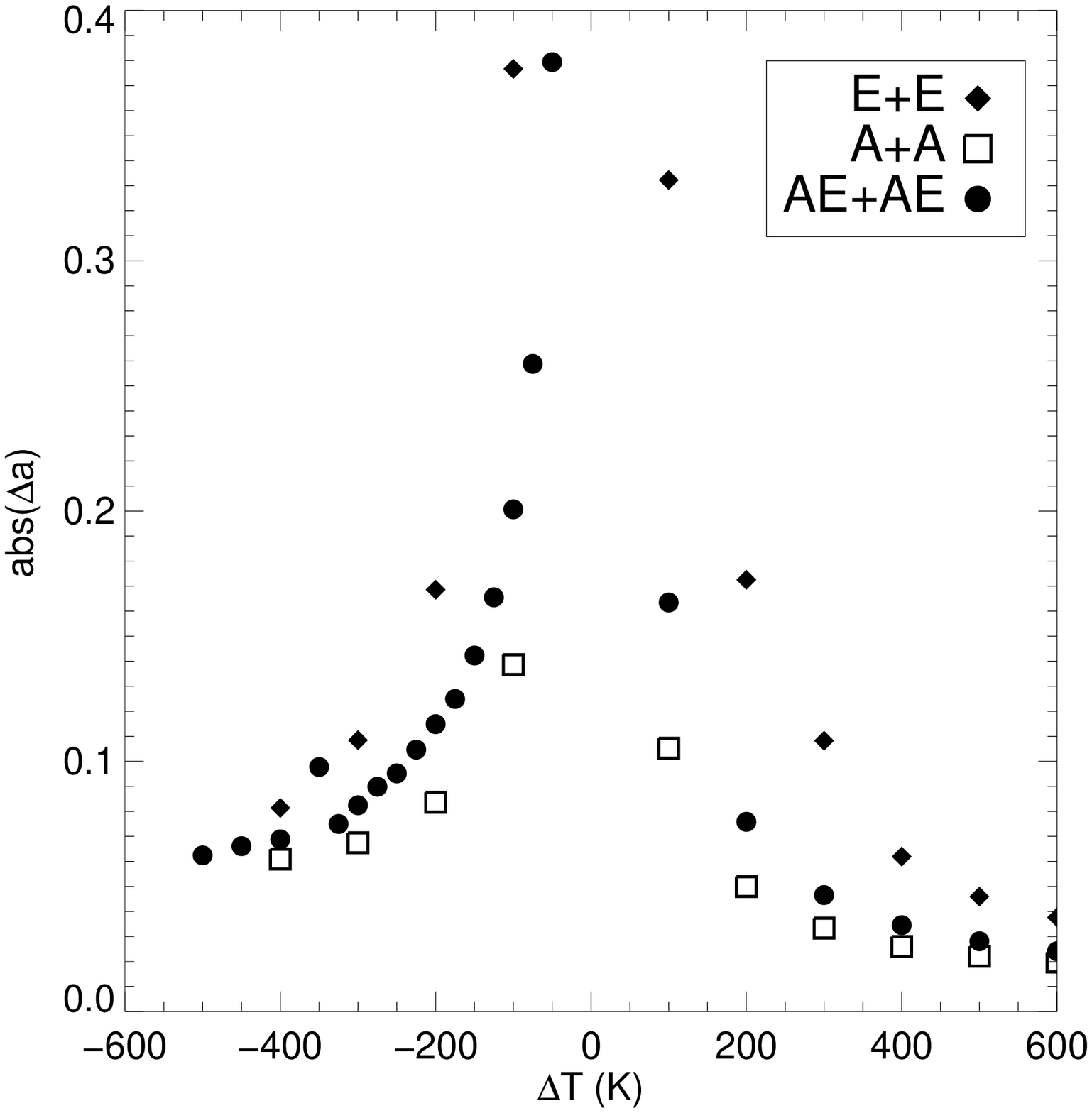} \\
\end{tabular}
\caption{top left and top right: predicted K$_{S}$ to J$_{S}$
  amplitude ratio $A_{K_S} / A_{J_S}$ as a function of $\Delta$T, the temperature
  difference between cloud components at T$_{1}$ and T$_{2}$, and for
  a filling fraction of the T$_{2}$ regions of 0.2.  Inhomogeneous cloud cover is plotted on the left (AE+E, A+E, A+AE)
  and homogeneous cloud cover is plotted on the right (E+E, A+A,
  AE+AE).  The yellow region denotes the values of the
  amplitude ratio that have previously been found for variable field
  brown dwarfs.
  Bottom left and bottom right: maximum change in filling fraction
  needed to produce the observed amplitude A$_{J_S}$ as a function of $\Delta$T.
\label{fig:modelcurves1}
}
\end{figure}

We then calculated synthetic fluxes in $J_{S}$ and $K_{S}$ for models
with all three cloud prescriptions, T$_\mathrm{eff}$ from 700-1700
K, and log(g)=4 (matching the measured log(g) of PSO J318.5-22
from Liu et al. 2013).  Then, considering different
values for $a$, we solved for $\Delta$a from Equation~\ref{eq:ff} 
for the maximum observed amplitude 
in $J_{S}$, with T$_{1}$ = 1000 K, different values of T$_{2}$, and varying
cloud prescriptions (plotted in the bottom 
panels of Fig.~\ref{fig:modelcurves1}
for a minimum T$_{2}$ filling fraction of 0.2).  
Filling fraction significantly varies for small
$\Delta$T, but only small variations in filling factor can
drive variability for abs($\Delta$T) $>$ 200 K.
Considering different
values for $a$, we calculated the variability amplitude ratio 
$A_{K_S} / A_{J_S}$ for the same combinations of
T$_{1}$, T$_{2}$, and varying cloud prescriptions.  We adopt the same 
convention as \citet[]{Rad12}, where the thicker cloud prescription 
is used for the F$_{1}$ regions.  In the inhomogenous cloud case, 
we also assume that the thinner
cloud producing the F$_{2}$ region is at a hotter 
T$_\mathrm{eff}$ than the F$_{1}$ regions (i.e. the thin 
cloud top is deeper in the atmosphere and thus hotter), so $\Delta$T = T$_{2}$ - T$_{1}$ $>$ 0.   
Representative results for predicted amplitude ratio are presented
in Fig.~\ref{fig:modelcurves1} -- similar to \citet[]{Rad12},
different minimum filling
fractions yield qualitatively similar results, so we 
present only $a$=0.2 results here.  Inhomogeneous
combinations of clouds are shown on the left, homogeneous
combinations on the right (i.e. hot spots instead of cloud patchiness
as the driver of variability).

Observations of variable brown dwarfs have generally found 
abs($A_{K_S} / A_{J_S}$) $<$ 1 \citep[see~e.g.~][]{Art09, Rad12,
  Rad14a, Wil14, Rad14b}, thus, we shade this region in yellow
in Fig. ~\ref{fig:modelcurves1}.  
As we have not yet covered a whole period of this variability 
nor do we have simultaneous multi-wavelength observations, 
we cannot determine $A_{K_S} / A_{J_S}$ with the data 
in hand.  It remains to be seen whether abs($A_{K_S} / A_{J_S}$) is
also $<$1 for PSO J318.5-22, which is much redder in $J-K$ than 
the high-g, bluer objects for which $A_{K_S} / A_{J_S}$ is robustly 
measured.  
Future observations that 
cover the entire period of variability at multiple wavelengths are
necessary to characterize the source of this variability.
However, in advance of these observations, it is instructive to
consider what amplitude ratios can be produced for young low surface
gravity objects with thick clouds.

In the case of inhomogeneous cloud cover (E+AE, E+A, A+AE), 
combinations of thick clouds can produce $A_{K_S} / A_{J_S} <$ 1, 
for $\Delta$T $>$150, similar to what was found by \citet[]{Rad12}
for the field early T 2MASS J21392676+0220226.
However, while \citet[]{Rad12} found that single component cloud models 
from \citet[]{Sau08} with f$_{sed}$=3 always have 
$A_{K_S} / A_{J_S} >$ 1, we do not find this to be the case 
with all of the \citet[]{Mad11} cloud models.  
This is true in the E+E case, but 
for combinations of thicker cloud models (AE+AE, A+A), 
 $A_{K_S} / A_{J_S} $ can be $<$1.  
Unlike \citet[]{Rad12}, who rule out homogeneous cloud 
cover with hot spots as a source of variability for the T1.5 brown dwarf
2MASS J21392676+0220226 based on a measured $A_{K_S} / A_{J_S} $ $<$1, 
a measurement of $A_{K_S} / A_{J_S} $ $<$1 for a young, low surface 
gravity objects with thick clouds would be consistent with both
inhomogeneous clouds (patchy cloud cover) and homogeneous clouds
(hot spots).

\section{Conclusions}
We detect significant variability in the young,
free-floating planetary mass object PSO J318.5-22, suggesting that
planetary companions to stars with similar colors (e.g. the HR 8799 planets) may also be
variable.  With variability amplitudes from 7-10$\%$ in J$_{S}$ at
two separate epochs over 3-5 hour observations, we constrain the period to 
$>$5 hours, likely $>$7-8 hours in the case of sinusoidal variation.  
In K$_{S}$, we marginally detect a variability trend of
up to 3$\%$ over our 3 hour observation.  Our marginal detection
suggests that the variability amplitude in K$_{S}$
may be smaller than that in J$_{S}$, but simultaneous
multi-wavelength observations are necessary to confirm this.  
Using the models of \citet[]{Mad11},
combinations of both homogeneous and inhomogeneous cloud prescriptions
can tentatively model variability with abs($A_{K_S} / A_{J_S}$) $<$ 1 
for young, low surface gravity objects with thick clouds.   

Only one exoplanet rotation period has been measured to date -- 7-9
hours for $\beta$ Pic b \citet[]{Sne14}.
PSO J318.5-22 is only the second young planetary mass object with
constraints placed on its rotational period and is likely also a fast rotator like $\beta$
Pic b, with possible rotation periods from
$\sim$5-20 hours.   PSO J318.5-22 is thus an important link
between the rotational properties of exoplanet companions
and those of old, isolated Y dwarfs with similar masses.

\acknowledgements
We thank the anonymous referee for useful comments which helped
improve this paper.  This work was supported by a consolidated grant from STFC.
E.B. was supported by the Swiss National Science Foundation (SNSF)
DH acknowledges support from the the ERC and DFG.

\end{document}